%% file: bare_jrnl.tex
\documentclass[journal]{IEEEtran}

\usepackage{cite}
\usepackage{orcidlink}

\usepackage{hyperref}
\usepackage{makeidx}
\usepackage{url}
\usepackage{graphicx}
\usepackage{enumitem}
\usepackage{amsmath}
\usepackage{amsfonts}
\usepackage{amssymb}
\usepackage{amsthm}
\usepackage{algorithmic}
\usepackage{booktabs}
\usepackage{tabularx}
\usepackage{makecell}
\usepackage{float}
\usepackage{mathtools}
\usepackage{caption}
\usepackage{subcaption}

\usepackage{bbm}

\hypersetup{hidelinks, colorlinks=true, citecolor=blue, linkcolor=blue, urlcolor=black}

\DeclareMathOperator{\tr}{tr}

\theoremstyle{definition}

\usepackage[ruled,vlined,linesnumbered]{algorithm2e}
\SetAlgorithmName{Algorithm}{Algorithm}{List of Algorithms}

\DeclareMathOperator*{\argmin}{arg\,min}

\usepackage{makecell}
\usepackage{multirow}

\begin{document}

\title{R-MTLLMF: Resilient Multi-Task Large Language Model Fusion at the Wireless Edge}

\author{
    Aladin Djuhera\IEEEauthorrefmark{1},
    Vlad~C.~Andrei\IEEEauthorrefmark{1},
    Mohsen Pourghasemian\IEEEauthorrefmark{2},
    Haris Gacanin\IEEEauthorrefmark{2},
    Holger~Boche\IEEEauthorrefmark{1},
    and~Walid~Saad\IEEEauthorrefmark{3} \\

    \IEEEauthorblockA{
        \IEEEauthorrefmark{1}Technical University of Munich, Munich, Germany,
        \IEEEauthorrefmark{2}RWTH Aachen University, Aachen, Germany, \\
        \IEEEauthorrefmark{3}Virginia Tech, Arlington, VA, USA\\
        Emails: \{aladin.djuhera, vlad.andrei, boche\}@tum.de, \{mohsen, harisg\}@dsp.rwth-aachen.de, walids@vt.edu
    }

\thanks{

\noindent 
The TUM group was supported by the german BMBF in the programs “Souver\"an. Digital. Vernetzt.” as part of the research hub 6G-life (Grant 16KISK002), QD-CamNetz (Grant
16KISQ077), QuaPhySI (Grant 16KIS1598K), and QUIET (Grant 16KISQ093), as well as by the project "6G Future Lab Bavaria". The RWTH Aachen group was supported by the 6GEM research hub under Grant 16KISK036K. W. Saad was supported by the U.S. National Science Foundation under Grant CNS-2114267.
}
}


\maketitle

\input{extras/abstract}

\IEEEpeerreviewmaketitle


\input{sections/section1/introduction}

\input{sections/section2/section2-A}
\input{sections/section2/section2-B}
\input{sections/section2/section2-C}

\input{sections/section3/section3-A}

\input{sections/section4/section4-A}

\input{extras/conclusion}

\ifCLASSOPTIONcaptionsoff
  \newpage
\fi

\bibliographystyle{IEEEtran}
\bibliography{IEEEabrv,bibliography}

\end{document}

%% file: extras/abstract.tex
\begin{abstract}
    \noindent
    Multi-task large language models (MTLLMs) are important for many applications at the wireless edge, where users demand specialized models to handle multiple tasks efficiently. However, training MTLLMs is complex and exhaustive, particularly when tasks are subject to change. Recently, the concept of model fusion via task vectors has emerged as an efficient approach for combining fine-tuning parameters to produce an MTLLM. In this paper, the problem of enabling edge users to collaboratively craft such MTLMs via tasks vectors is studied, under the assumption of worst-case adversarial attacks. To this end, first the influence of adversarial noise to multi-task model fusion is investigated and a relationship between the so-called weight disentanglement error and the mean squared error (MSE) is derived. Using hypothesis testing, it is directly shown that the MSE increases interference between task vectors, thereby rendering model fusion ineffective. Then, a novel resilient MTLLM fusion (R-MTLLMF) is proposed, which leverages insights about the LLM architecture and fine-tuning process to safeguard task vector aggregation under adversarial noise by realigning the MTLLM. The proposed R-MTLLMF is then compared for both worst-case and ideal transmission scenarios to study the impact of the wireless channel. Extensive model fusion experiments with vision LLMs demonstrate R-MTLLMF's effectiveness, achieving close-to-baseline performance across eight different tasks in ideal noise scenarios and significantly outperforming unprotected model fusion in worst-case scenarios. The results further advocate for additional physical layer protection for a holistic approach to resilience, from both a wireless and LLM perspective.
\end{abstract}


%% file: sections/section1/introduction.tex
\section{Introduction and Motivation}
Future 6G wireless networks will enable reliable, low-latency, and intelligent communications to support critical technologies such as artificial intelligence (AI) \cite{saad2024artificial}, distributed and collaborative machine learning (ML) \cite{chen2021distributed}, and intelligent edge computing \cite{zhu2023pushing}. Recent advancements in large language models (LLMs) in particular have motivated several use cases for the wireless industry, including network design, resource allocation, and standardization \cite{xu2024largemultimodalmodelslmms}. In this generative AI vision of the future, distributed agents will play a central role in fine-tuning, sharing and merging LLMs at the wireless edge. For example, \cite{GenAINet} presents a framework where distributed agents use LLMs for reasoning and collaboration in wireless networks, supporting knowledge-based edge device queries.

However, due to the large size of LLMs, traditional federated learning (FL) \cite{mcmahan2023communicationefficient} approaches are often inefficient in scenarios where users demand multi-task capabilities that may change over time. FL typically requires extensive training over multiple communication rounds and assumes that users have limited data and compute resources, which can be restrictive for specialized edge devices with significant processing power and data. In contrast, multi-task model fusion (MTMF) enables each client to contribute a task-specific model directly without collaboratively fine-tuning a single multi-task LLM (MTLLM). Instead, the authors in \cite{ilharco2023editingmodelstaskarithmetic} introduce the concept of \textit{task vectors} which capture task-specific adjustments that can be efficiently aggregated to produce a high-performing MTLLM with minimal computation and communication overhead at the aggregator. This approach avoids the inflexibility of FL by enabling quick adaptation to new tasks as they arise, without re-training from scratch. In addition, it is amenable for various \textit{parameter-efficient fine-tuning} (PEFT) methods \cite{tang2024parameterefficientmultitaskmodel}. Thus, MTMF is better suited for scenarios where user requirements and tasks change dynamically, providing a more efficient and adaptable framework than conventional FL.

However, task vector transmissions may be prone to both adversarial attacks and suboptimal channel conditions, which may significantly increase task interference. In our previous work in \cite{djuhera2024rsfllmjammingresilientframework}, for example, we investigated worst-case jamming for split FL where intermediate model parameters were transmitted. We explicitly showed that worst-case jammers lead to worst-case performance and developed an effective, sensing-assisted physical layer protection scheme. In contrast, MTMF represents a \textit{distributed inference} scenario in which adversarial noise during task vector aggregation may be catastrophic. 

The main contribution of this paper is, thus, an analysis of adversarial attacks on MTMF over wireless channels and the development of a resilient MTLLM fusion framework (R-MTLLMF), which leverages the unique architecture and fine-tuning characteristics of LLMs within MTMF to safeguard task vector aggregation against adversarial noise. To this end, we investigate worst-case channel conditions in which adversarial noise, induced by a malicious attacker with optimal attack strategy, considerably decreases the communication rate below a certain threshold, such that reactive network defenses and other upper-layer resilience mechanisms are insufficient \cite{jorswieck_performance_2004}, thus requiring a resilient physical layer solution \cite{andrei_resilient}. We take this special case as an example to study the worst-case impact of adversarial noise to MTMF, and to understand in how far AI-based resilience mechanisms alone can be effective without the help of additional physical layer protection schemes. In summary, our key contributions include:
\begin{itemize}[leftmargin=*]
    \item We first introduce the framework of MTMF via task vectors at the wireless edge as a flexible multi-task alternative to FL.
    \item Then, we derive an explicit relationship between the so-called weight disentanglement error (WDE), which characterizes the cross-task interference, and the wireless mean squared error (MSE). We show that high MSE increases the task interference, rendering MTMF inefficient. We further study the worst-case noise covariance for both the sum rate and the strongest user in multiple-input multiple-output (MIMO) channels, as a benchmark for the most adverse channel noise conditions, induced by the worst-case attacker.
    \item Next, we introduce R-MTLLMF, which safeguards task vector aggregation by realigning the MTLLM using few-shot fine-tuning and parameter freezing. We also discuss the importance of resilient physical layer schemes to further enhance our AI-based framework for extreme conditions.
    \item Finally, we conduct extensive simulations with vision transformer LLMs (ViT-LLMs). We show that worst-case noise results in worst-case multi- and single-task performance across eight different datasets. In contrast, R-MTLLMF achieves excellent performance under ideal, and strong performance in worst-case scenarios, thus enhancing the resilience of MTLLMF using only AI methods. Nonetheless, additional wireless countermeasures may still be necessary.
\end{itemize} 

The rest of this paper is organized as follows. The MTMF system model, relationship between WDE and MSE, and design of the worst-case noise covariance are given in Section II. R-MTLLMF is presented in Section III. Section IV discusses simulation results, and conclusions are drawn in Section V.

%% file: sections/section2/section2-A.tex
\section{Multi-Task Model Fusion at the Wireless Edge}
\begin{figure}[!t]
    \centering
    \includegraphics[width=0.9\linewidth]{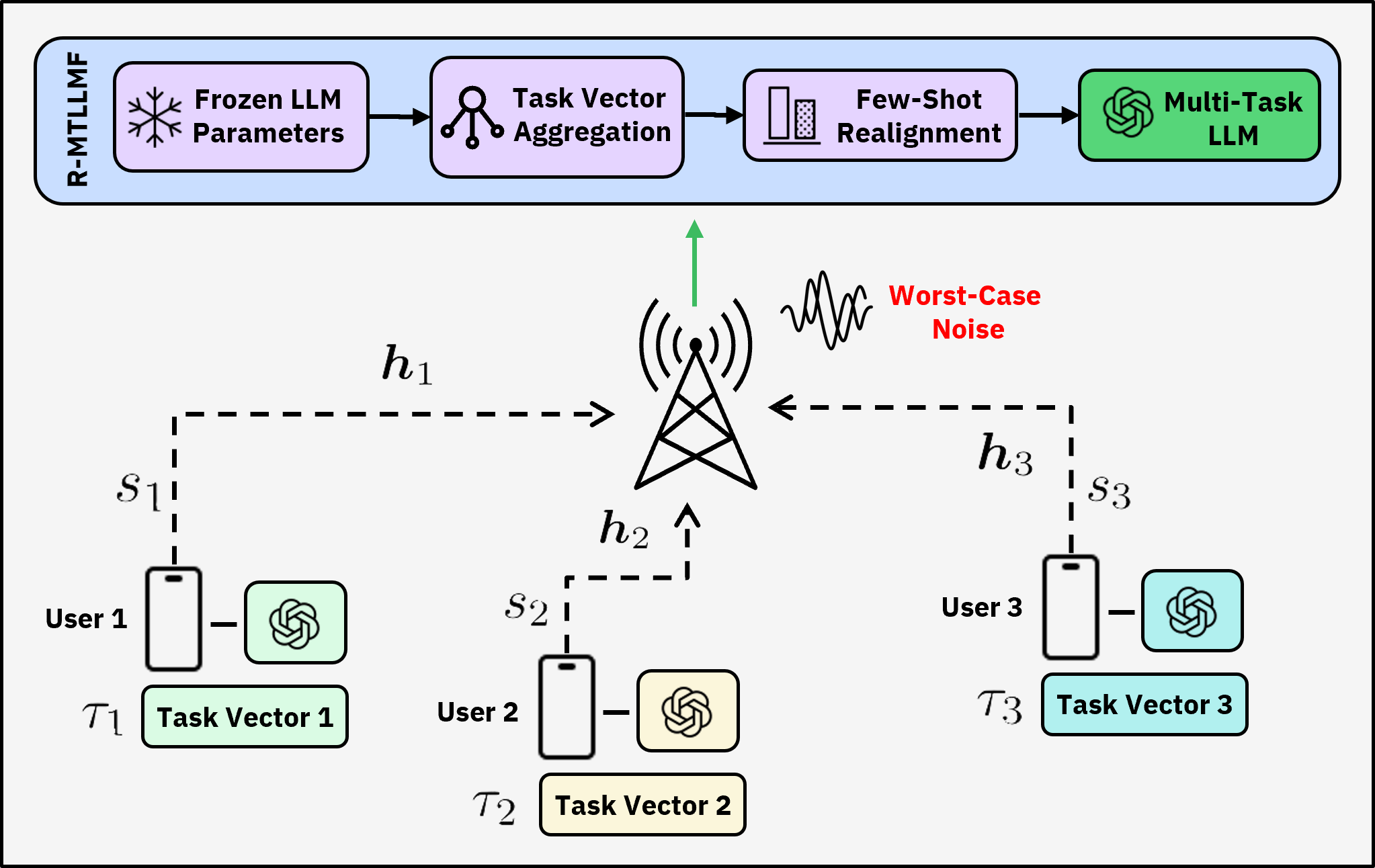}
    \caption{MTMF system model where users collaboratively construct a MTLLM. R-MTLLMF safeguards task vector aggregation by realigning the perturbed model and freezing sensitive LLM parameters that do not change significantly.}
    \vspace{-0.5cm}
    \label{fig:system_model}
\end{figure}

In this section, we introduce MTMF, derive the relationship between the WDE and the wireless MSE, and design the worst-case adversarial noise covariance. The system model as well as our R-MTLLMF framework to safeguard MTMF at the wireless edge are shown in Figure \ref{fig:system_model}. 

\subsection{Wireless MTMF System Model}
We consider a collaborative inference setup in which a set $\mathcal{Q}$ of $Q$ single-antenna users construct MTLLMs for several different tasks at a time, starting from the same pre-trained LLM. To this end, each user $q \in \mathcal{Q}$ first fine-tunes the LLM on its respective task and data, and computes the task vector $\tau_q$ by subtracting fine-tuned and base model parameters, i.e.
\begin{align}
    \tau_q = \theta_{\text{fine-tuned}, q} - \theta_{\text{base}} \ .
\end{align}

Then, each user encodes its task vector and transmits unit variance symbols $s_q$ with power $p_q \leq P_q$ over a MIMO multiple access channel (MAC) to a base station (BS) with $N_R$ receive antennas. The wireless transmission model is given as: 

\newpage
\noindent
\begin{align}
    \boldsymbol{y} = \sum_{q=1}^{Q} \boldsymbol{h}_q \sqrt{p_q}s_q + \boldsymbol{z} \ , \ \boldsymbol{z} \sim \mathcal{N} (\boldsymbol{0}, \boldsymbol{C}_{z}) \ ,
\end{align}
with noise covariance $\boldsymbol{C}_{z}$, total noise power $P_{N} \geq \tr{ ( \boldsymbol{C}_{z} ) }$, and with Rayleigh distributed user channels $\boldsymbol{h}_q$. We further assume that the BS performs successive interference cancellation (SIC), and set the SIC ordering to $p_{1} \|  {\boldsymbol{h}_1} \|_2^2  \leq\dots\leq p_{Q} \| {\boldsymbol{h}_Q} \|_2^2$. With that, the achievable rate at each user $q$ will be given by:
\begin{align}
    R_q &= \log\left(1 + p_q \boldsymbol{h}_q^H \boldsymbol{X}_{q}^{-1} \boldsymbol{h}_q\right) ,
\end{align}
where we abbreviated $\boldsymbol{X}_q = \sum_{q^\prime > q} p_{q^{\prime}} \boldsymbol{h}_{q^\prime} \boldsymbol{h}_{q^\prime}^H + \boldsymbol{C}_{z}$. Further, we can write the corresponding sum rate for all users as:
\begin{align}
    R = \log\left(1 + \sum_{q=1}^{Q} p_q \boldsymbol{h}_q^H \boldsymbol{C}_{z}^{-1} \boldsymbol{h}_q \right).
    \label{eq:sum_rate}
\end{align}

We assume that the BS performs minimum MSE (MMSE) equalization, such that the achievable MSE per user $q$ is 
\begin{align}
    \mu_q = \left(1 + p_q \boldsymbol{h}_q^H \boldsymbol{X}_{q}^{-1} \boldsymbol{h}_q \right)^{-1} = \mathrm{e}^{-R_q} .
    \label{eq:mse}
\end{align}

Upon receiving the task vector symbols $\hat{s}_q$, the BS then aggregates all decoded $\hat{\tau}_q$ to define the new MTLLM parameters using a heuritsic scaling term $\lambda_N \in [0, 1]$, which is only dependent on the number of tasks $N$ as outlined in \cite{ilharco2023editingmodelstaskarithmetic}, i.e.
\begin{align}
    \theta_{\text{MTLLM}} = \theta_{\text{base}} + \lambda_N \sum_{q=1}^{N=Q} \hat{\tau}_i .
    \label{eq:aggregation}
\end{align}

The BS then broadcasts $\theta_{\text{MTLLM}}$ to the requesting parties for inference and thus acts as an on-demand model fusion server.

Such individually fine-tuned LLMs, initialized from the same pre-trained model $\theta_{\text{base}}$, effectively share a part of the optimization trajectory and can therefore often be efficiently merged while increasing the cross-task performance \cite{ilharco2023editingmodelstaskarithmetic}. However, worst-case channel conditions may lead to decoding errors where $\hat{\tau}_q \neq \tau_q$, thereby significantly altering the decoded task vectors and introducing severe task interference. In order to analyze the effect of such errors on MTMF, we investigate worst-case adversarial noise attacks \cite{jorswieck_performance_2004} and develop an AI-based protection scheme using insights on LLM fine-tuning. 

%% file: sections/section2/section2-B.tex
\subsection{Wireless Influence on Cross-Task Interference}
In general, MTMF may introduce task interference if orthogonality between task vectors cannot be ensured, particularly as the task count increases \cite{ilharco2023editingmodelstaskarithmetic}. The WDE \cite{tang2024parameterefficientmultitaskmodel} measures cross-task interference by comparing model predictions for individual versus simultaneous task vector additions. For $N > 2$ tasks and task dataset distribution $P(D{\tau_i})$, it is defined as: 
\begin{align}
    \xi &(\lambda) = \sum_{i=1}^{N} \mathbb{E}_{x \sim P(D{\tau_i})} \Gamma \ ,
    \label{eq:WDE}
\end{align}
where $\Gamma$ is a distance metric, for example the indicator function $\mathbbm{1} (.)$ for classification tasks, between two model outputs $f(x, \theta)$, with input data $x$ and model parameters $\theta$, i.e. 
\begin{align} 
    &\Gamma = \left[ \mathbbm{1} \left(f(x; \theta_{\text{base}} + \lambda_i \tau_i), f \left(x; \theta_{\text{base}} + \lambda_N \sum_{j=1}^{N} \tau_j\right)\right) \right] .
    \label{eq:Gamma}
\end{align} 

A lower value of $\xi$ indicates that the task vectors are well-disentangled, implying less task interference. In order to better capture the change when task vectors are exposed to wireless noise, we model the indicator function as a hypothesis test:
\begin{itemize}[leftmargin=*]
    \item \textbf{Null Hypothesis} $H_0$: The noise in the task vectors does not significantly alter the model's predictions. 
    \item \textbf{Alternative Hypothesis} $H_1$: The noise in the task vectors leads to a significant change in the model's predictions.
\end{itemize}

Let $z_u(x)$ and $z_d(x)$ be the \textit{logits} of the undisturbed and disturbed MTLLMs, respectively. The test statistic is the ratio $R(x) = \frac{z_u(x)}{z_d(x)}$. Under $H_0$, we thus require $R(x) \approx 1$. In case of noisy $\hat{\tau}_q$, the difference in model parameters will be:
\begin{align} 
    \Delta \theta = \sum_{q} \lambda_N \underbrace{(\tau_q + \Tilde{\epsilon}_q)}_{\hat{\tau}_q}  - \sum_{q} \lambda_N \tau_q = \sum_{q} \epsilon_q, 
\end{align}
where $\epsilon_q = \lambda_N \Tilde{\epsilon}_q$ reflects the cumulative post-decoding errors. Assuming that $\epsilon_q$ are small, we can perform a first-order Taylor expansion of the logits around the undisturbed parameters, i.e. 
\begin{align} 
    z_d(x) \approx z_u(x) + J_{\theta} \Delta \theta = z_u(x) + J_{\theta} \sum_{q} \epsilon_q, 
\end{align}
where $J_{\theta}$ is the Jacobian of the logits with respect to $\theta$. The deviation in the logits is thus directly proportional to $\epsilon_q$. With $J_{\theta} \sum_{q} \epsilon_q \ll z_u(x)$, we then have for the test statistic that $R(x) \approx 1 - \frac{J_{\phi} \sum_{q} \epsilon_q}{z_u(x)}$, which shows that $|R(x) - 1|$ is proportional to $\epsilon_q$. For both to remain small, we thus minimize the wireless sum MSE, i.e. $\sum_{q} \mu_q = \sum_{q} \mathbb{E} \left[ |\Tilde{\epsilon}_q|^2 \right]$. To determine the hypothesis threshold $T$, we further consider the distribution of $R(x)$ under $H_0$. Assuming that the perturbations $\epsilon_q$ are random variables with zero mean and variance related to the MSE, $|R(x) - 1|$ can be modeled as:
\begin{align} 
    \text{Var}[|R(x) - 1|] \approx \left( \frac{J_{\phi}}{z_u(x)} \right)^2 \sum_{q} \mu_q \ .
\end{align}
We thus set $T$ based on the standard deviation of $|R(x) - 1|$, i.e. $T = z_{\beta} \cdot \sqrt{\text{Var}[|R(x) - 1|]}$, where $z_{\beta}$ is the z-score w.r.t. to a significance level $\beta$. Alternatively, $T$ can also be chosen depending on additional insights, for example, on the model architecture and data quality. The null hypothesis is then rejected if $|R(x) - 1| > T$. Thus, minimizing the sum MSE increases the likelihood of satisfying $H_0$, i.e.
\begin{align} 
    \Pr(|R(x) - 1| \leq T) \rightarrow 1 \quad \text{as} \quad \sum_{q} \mu_q \rightarrow 0 \ .
\end{align}

Modeling the indicator function in \eqref{eq:Gamma} as a hypothesis test reveals that the wireless MSE directly impacts the cross-task interference. When \( H_0 \) is unmet, interference increases with \( \Gamma \) and WDE values, degrading multi-task performance due to increased weight entanglement. This insight underscores the need for resilient MTMF strategies under worst-case adversarial attacks where standard resilience measures fall short. As we explore AI-based methods for mitigating adversarial noise, understanding the limits under such conditions is essential before deploying additional wireless or AI countermeasures.

%% file: sections/section2/section2-C.tex
\subsection{Worst-Case Adversarial Noise Covariance Design}
To design $\boldsymbol{C}_{z}$ for the worst-case attack strategy, we consider adversarial noise for both system sum rate \eqref{eq:P1} and strongest user \eqref{eq:P2}. This allows us to study the impact of the attack on the network as well as on individual task vectors. Formally, we investigate the following saddle point problems:

\begin{align}
    \boldsymbol{C}_{z}^{(1)} &= \argmin_{\boldsymbol{C}_{z} \in \mathcal{S}(P_{N})} \max_{\boldsymbol{p} \in \mathcal{P}}  R(\boldsymbol{p}, \boldsymbol{C}_{z}) , \tag{P1}\label{eq:P1}\\
    \boldsymbol{C}_{z}^{(2)} &= \argmin_{\boldsymbol{C}_{z} \in \mathcal{S}(P_{N})} \max_{\boldsymbol{p}\in \mathcal{P}} \max_{q \in \mathcal{Q}} R_{q}(\boldsymbol{p}, \boldsymbol{C}_{z}) , \tag{P2}\label{eq:P2}
\end{align}
where $\boldsymbol{p} = [p_1, \dots, p_{Q}]\in\mathbb{R}^Q$, $\mathcal{P} = \{\boldsymbol{p} \,|\, \boldsymbol{p} \geq \boldsymbol{0},\, p_{q} \leq P_{q}\}$, and $\mathcal{S}(P_{N}) = \{\boldsymbol{C}_{z} \, | \, \boldsymbol{C}_{z}=\boldsymbol{C}_{z}^H, \, \boldsymbol{C}_{z} \succcurlyeq \boldsymbol{0}, \, \tr{(\boldsymbol{C}_{z})}\leq P_{N}\}$.

\subsubsection{Ideal Case}
First, we discuss the ideal transmission case corresponding to the scenario in which we solve the inner maximizations in \eqref{eq:P1} and \eqref{eq:P2}. In this setup, \( \mathbf{h}_q^H \mathbf{C}_{z}^{-1} \mathbf{h}_q \geq 0 \), making the sum rate an increasing function of \( \mathbf{p} \). Since \( \mathcal{P} \) is a convex set, the maximum in \eqref{eq:P1} and \eqref{eq:P2} is attained at the boundary, i.e., when each user transmits with maximum power \( p^{*}_{q} = P_{q} \). We define this as the ideal case, used as a benchmark for typical channel conditions in our experiments. 

\subsubsection{Solution to \eqref{eq:P1}}
Setting \( p^{*}_{q} = P_{q} \) accordingly and defining \( \mathbf{P} = \text{diag}(\mathbf{p}^{*})\in\mathbb{R}^{Q\times Q} \), we can rewrite \eqref{eq:P1} as:
\begin{align}
    \boldsymbol{C}_{z}^{(1)} = \min_{\boldsymbol{C}_{z}\in \mathcal{S}(P_{N})} \log\det\left(\boldsymbol{\mathrm{I}} + \boldsymbol{H}\boldsymbol{PH}^H\boldsymbol{C}_{z}^{-1} \right) .
    \label{eq:P1_rewritten}
\end{align}
By Hadamard's inequality, \( \mathbf{C}_{z} = \mathbf{U}\mathbf{\Sigma}\mathbf{U}^H \) is the optimal solution with eigenvectors \( \mathbf{U} \) of \( \mathbf{H}\mathbf{PH}^H \), reducing the problem to optimizing the eigenvalues \( \mathbf{\Sigma} = \text{diag}(\{\sigma_{i}\}_{i=1}^{N_{R}}) \) as in \cite{jorswieck_performance_2004}:
\begin{align}
    \sigma_{i} = \frac{P_i \upsilon_i}{2}\left(\sqrt{1 + \frac{4}{P_i \upsilon_i \nu}} - 1 \right),
\end{align}
where \( \upsilon_i \) are the eigenvalues of \( \mathbf{H}\mathbf{PH}^H \), and \( \nu \geq 0 \) is chosen to satisfy \( \sum_{i=1}^{N_R} \sigma_i \leq P_N \), computed via the bisection method.

\subsubsection{Solution to \eqref{eq:P2}}
Following similar arguments for the inner maximization, we use the SIC ordering from Section II.A, noting that \( \max_{q\in\mathcal{Q}}R_q = R_{Q} \). This rewrites \eqref{eq:P2} as:
\begin{align}
    \mathbf{C}_{z}^{(2)} = \min_{\mathbf{C}_{z}\in \mathcal{S}(P_{N})} \log\det\left(\mathbf{\mathrm{I}} + P_{Q}\mathbf{h}_{Q}\mathbf{h}_{Q}^H\mathbf{C}_{z}^{-1} \right),
    \label{eq:P2_rewritten}
\end{align}
where the optimal \( \mathbf{C}_{z} = P_{N} \mathbf{h}_{Q}\mathbf{h}_{Q}^H \Big/ \| {\mathbf{h}_{Q}} \|_2^2 \) is obtained by aligning the noise covariance with the strongest user channel.

\newpage
Problems \eqref{eq:P1} and \eqref{eq:P2} represent worst-case scenarios where an adversarial attacker strategically injects noise, which renders common resilience mechanisms like ARQ ineffective \cite{jorswieck_performance_2004}. Given the vulnerability of task vectors to adversarial noise in MTMF, we next address the challenge on how to design an MTMF framework that remains resilient under such adversarial attacks without relying solely on conventional physical layer resilience mechanisms. In the next section, we introduce our proposed R-MTLLMF, which employs AI-driven methods to counter adversarial noise in MTMF.

%% file: sections/section3/section3-A.tex
\section{R-MTLLMF: AI-Driven Resilience Against Adversarial Noise}
To address the challenges posed by adversarial noise on task vector aggregation in MTMF, we approach resilience from an AI perspective by introducing R-MTLLMF, an AI-based framework designed to mitigate adversarial cross-task interference. As shown in Figure \ref{fig:system_model}, it consists of two core modules which work together by leveraging insights on both LLM architecture and task vector characteristics in MTMF:
\begin{itemize}[leftmargin=*]
    \item \textbf{Task Vector Aggregation with Frozen LLM Parameters}: Given the severity of noise at inference time, we freeze sensitive position, patch and class embeddings of the base model $\theta_{\text{base}}$, and reload them into $\theta_{\text{MTLLM}}$ after task vector aggregation. Freezing these parameters ensures that fundamental data representations remain intact, reducing the risk of noise-induced shifts in essential encodings. This is a valid approach since fine-tuning pre-trained models mostly affects the self-attention layers, and since $\theta_{\text{base}}$ remains identical across tasks \cite{ilharco2023editingmodelstaskarithmetic, tang2024parameterefficientmultitaskmodel}. Thus, by preserving embeddings, we maintain the integrity of structural representations learned during pre-training, thereby stabilizing the model’s output.

    \item \textbf{Few-Shot Realignment}: To correct for noise-induced perturbations in self-attention parameters, we apply a few-shot fine-tuning step to realign the MTLLM on a small, representative data subset (typically 10 samples per class). This can be sourced from episodic memory, as in \cite{chitale2023taskarithmeticloracontinual}, or a public dataset for privacy considerations, and can be performed at the MTMF BS. By realigning the attention mechanism on few samples, the model restores the contextual relationships between tokens that may be distorted due to noise. Drawing on insights from task vector research in \cite{ilharco2023editingmodelstaskarithmetic}, this heuristic relies on the notion that noise, if not substantial, leaves the underlying task vector directionality of $\hat{\tau}_q$ largely intact, allowing the model’s performance to be recovered with minimal fine-tuning. Thus, realignment addresses moderate perturbations, restoring the model's reasoning capabilities.
\end{itemize}

R-MTLLMF introduces an AI-based protection scheme to ensure resilient model outputs at inference time. It performs few-shot fine-tuning to realign the perturbed model, and freezes sensitive embedding parameters to ensure critical integrity of data representations. While typical resilience methods in distributed learning often rely on additional physical layer measures \cite{djuhera2024rsfllmjammingresilientframework}, R-MTLLMF focuses on enhancing resilience within the LLM architecture itself. This heuristic, grounded in recent findings on task vectors, thus provides a flexible, low-overhead resilience layer, optimized for dynamic MTMF scenarios. However, R-MTLLMF may be insufficient if the channel conditions are detrimental, such that the low-noise assumption for few-shot realignment is invalidated. To this end, we will investigate both ideal and worst-case adversarial noise scenarios in the subsequent experiments, demonstrating R-MTLLMF's effectiveness in typical conditions while examining its limits under adversarial interference. To ensure the best possible protection against worst-case conditions, including noise and adversarial attacks, R-MTLLMF can be further augmented by physical layer resilience schemes as in \cite{andrei_resilient}. However, this is out of the scope of this study.

%% file: sections/section4/section4-A.tex
\section{Simulation Results and Analysis}
We follow the experimental setup in \cite{tang2024parameterefficientmultitaskmodel} and fine-tune OpenAI's CLIP ViT-B/16 ViT-LLM \cite{CLIP} for 2000 iterations with batch size 128 on eight different image classification datasets (tasks): MNIST, Cars \cite{cars}, DTD \cite{dtd}, EuroSAT \cite{eurosat}, GTSRB \cite{gtsrb}, RESISC45 \cite{resisc}, SUN397 \cite{sun}, and SVHN \cite{svhn}. We then compute the task vectors and construct the MTLLM with parameters $\theta_{\text{MTLLM}}$ as defined in \eqref{eq:aggregation} with heuristic values for $\lambda_N$ from \cite{ilharco2023editingmodelstaskarithmetic}. We vary the number of tasks between 2 and 8, and, for each case, we generate all possible task combinations and evaluate the MTMF cross-task performance by measuring accuracy on each dataset, normalized to the individual fine-tuning performance. We report the average of the normalized accuracies as in \cite{tang2024parameterefficientmultitaskmodel}. To simulate ideal and adversarial channel noise conditions for sum rate (SR) and strongest user (SU) per \eqref{eq:P1} and \eqref{eq:P2}, we consider the MIMO setup from Section II with $|\mathcal{Q}| = 8$ single-antenna users, each of which fine-tunes on one of above datasets, and $N_R = 16$ BS receive antennas. We set the maximum transmit power to 0.1 mW. We initialize the user positions randomly between 100 to 1000 meters from the BS and employ 64-QAM modulation. To induce parameter deviation from post-decoding errors $\Tilde{e}_q$, we compute the MSE and add Gaussian noise with variance proportional to it to the task vectors, resulting in $\hat{\tau}_q = \tau_q + \Tilde{\epsilon}_q$. We apply R-MTLLMF with few-shot realignment, using 10 random samples per class. 

\subsubsection{R-MTLLMF Results}
\begin{figure}[htbp]
    \centering
    \includegraphics[width=0.47\textwidth]{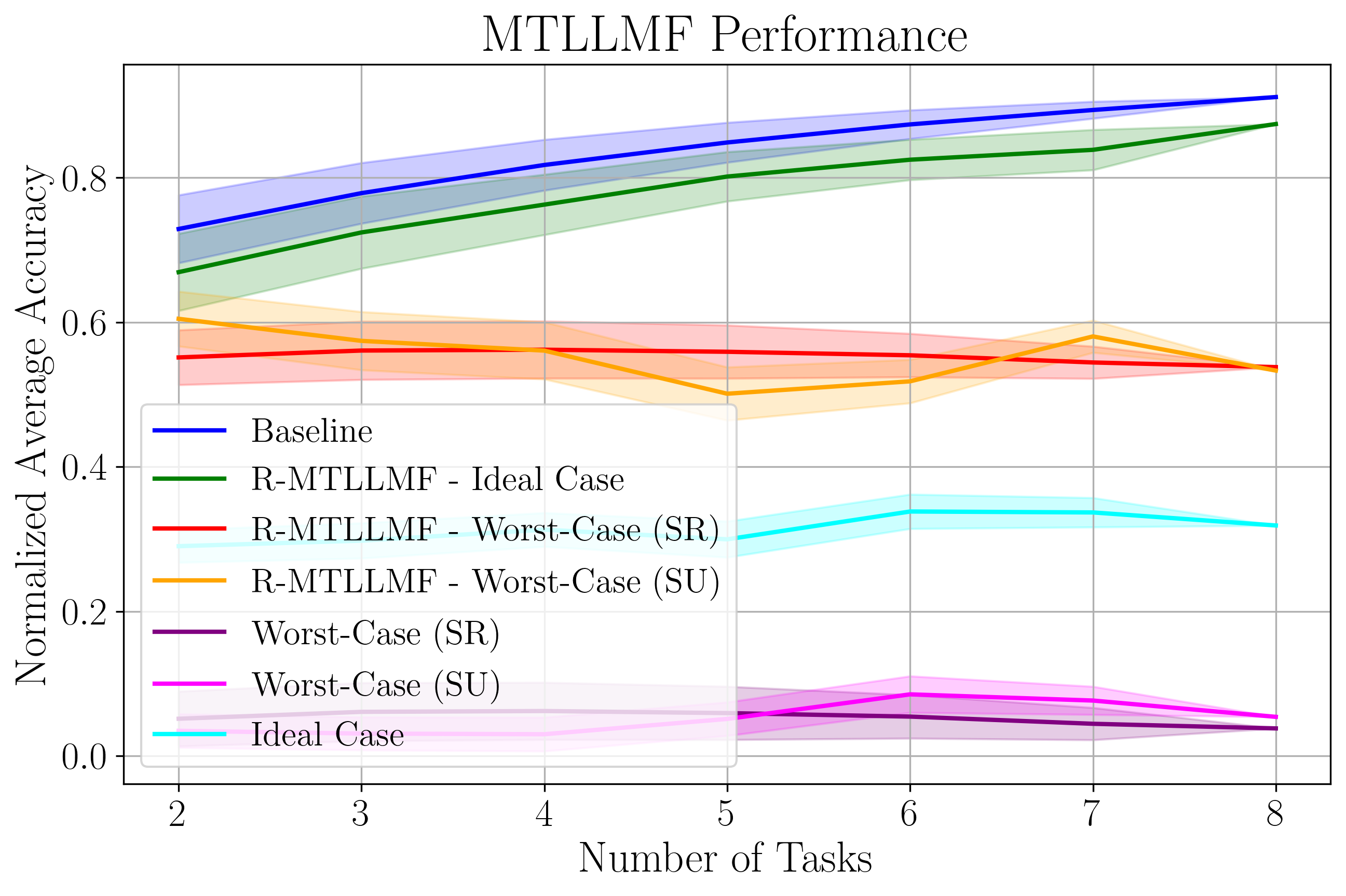}
    \caption{Results for MTMF with 8 tasks. R-MTLLMF achieves close-to-baseline performance for the ideal transmission case, while having degraded performance under worst-case noise for sum rate (SR) and strongest user (SU).}
    \label{fig:mtllmf_performance}
\end{figure}

Figure \ref{fig:mtllmf_performance} shows results for 8 tasks where shaded areas represent the variance in accuracies for all task combinations. R-MTLLMF achieves close-to-baseline performance for ideal transmission, having 89.1\% normalized average accuracy compared to the 91.2\% baseline when aggregating all 8 tasks. In contrast, R-MTLLMF has equally degraded performance for both SR and SU worst-case scenarios with 54.8\% and 53.3\% accuracies, respectively, due to the significantly increased noise during task aggregation. Whereas the ideal scenario results in a signal-to-noise-ratio (SNR) of $-16.2$ dB with an average MSE of $0.013$ across all 8 users, results for SU (SR) yield an SNR of -21.7 dB (-27.3 dB) with an average MSE of $0.16$ ($0.21$). The severe increase in noise shows some limitations of our AI-based framework where neither parameter freezing nor few-shot realignment can fully restore the model's classification capabilities as the cross-task interference is too high. However, even in this worst-case, R-MTLLMF ensures significantly better model performance (9-14 times better) as compared to scenarios without protection. More concretely, unprotected SR and SU scenarios lead to catastrophic perturbances with accuracies between 4\% and 6\%, 

\newpage
\noindent
resulting in worst-case single- and multi-task performance, which ultimately renders MTMF ineffective. Surprisingly, even the unprotected ideal case results in subpar performance with total aggregation accuracy of 36.3\% due to noise at inference time. Furthermore, accuracies seem to be generally non-increasing with the number of tasks, except for R-MTLLMF in the ideal case. This indicates that the perturbed model is unable to generalize across tasks. In addition, the heuristic choice for $\lambda_N$ may be suboptimal, indicated by random improvements and deteriorations. This signifies the importance of resilient aggregation strategies in distributed inference scenarios, where noise, even if small, may be detrimental.   

\subsubsection{Weight Disentanglement Error Analysis}
\begin{figure}[htbp]
    \centering
    \includegraphics[width=0.47\textwidth]{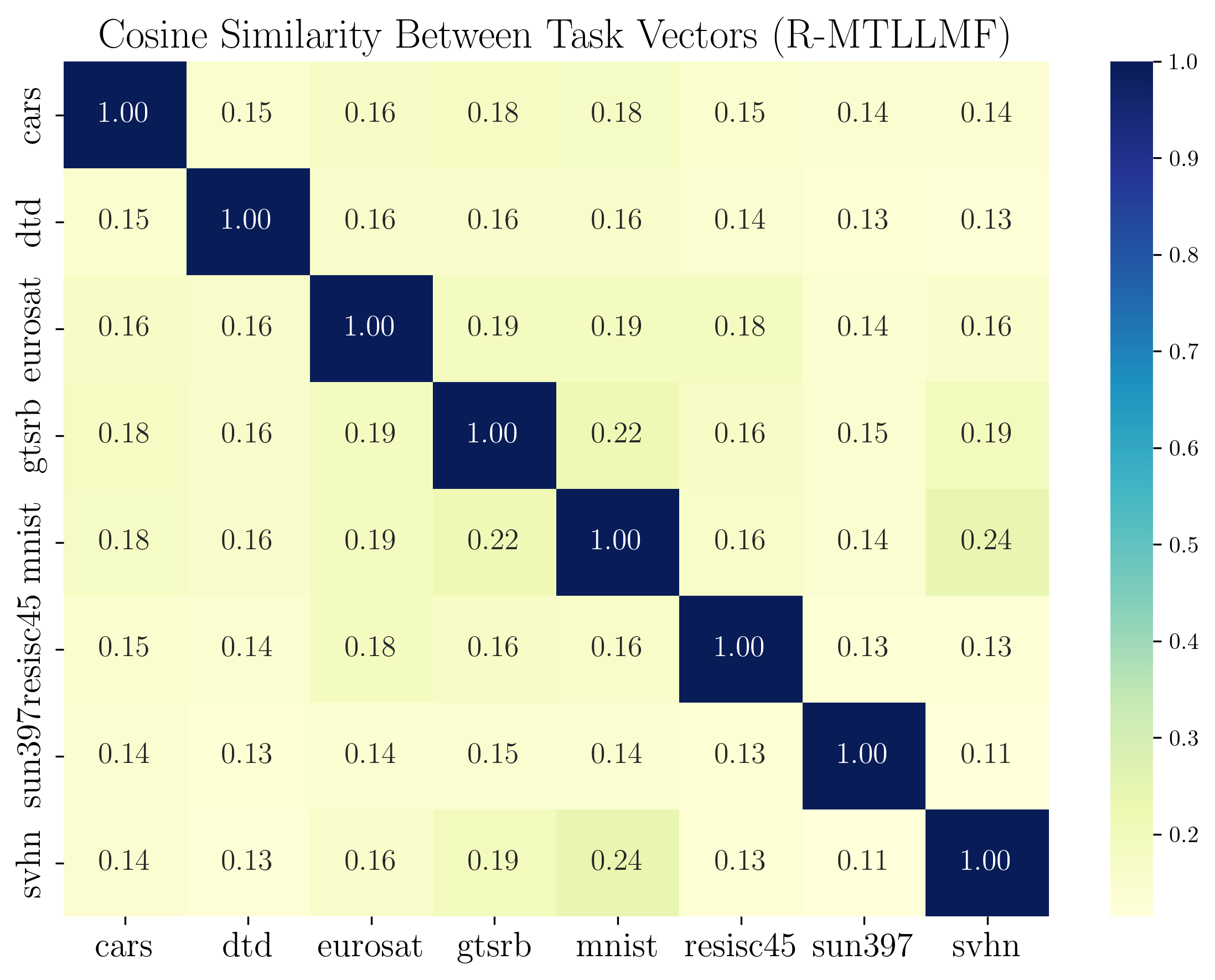}
    \caption{Cosine similarities between task vectors for R-MTLLMF (ideal case) as a measure for weight disentanglement. All $\hat{\tau}_q$ have small similarity scores, indicating less cross-task interference and increased orthogonality.}
    \label{fig:mtllmf_wde_RMTLLMF}
\end{figure}

\begin{figure}[htbp]
    \centering
    \includegraphics[width=0.47\textwidth]{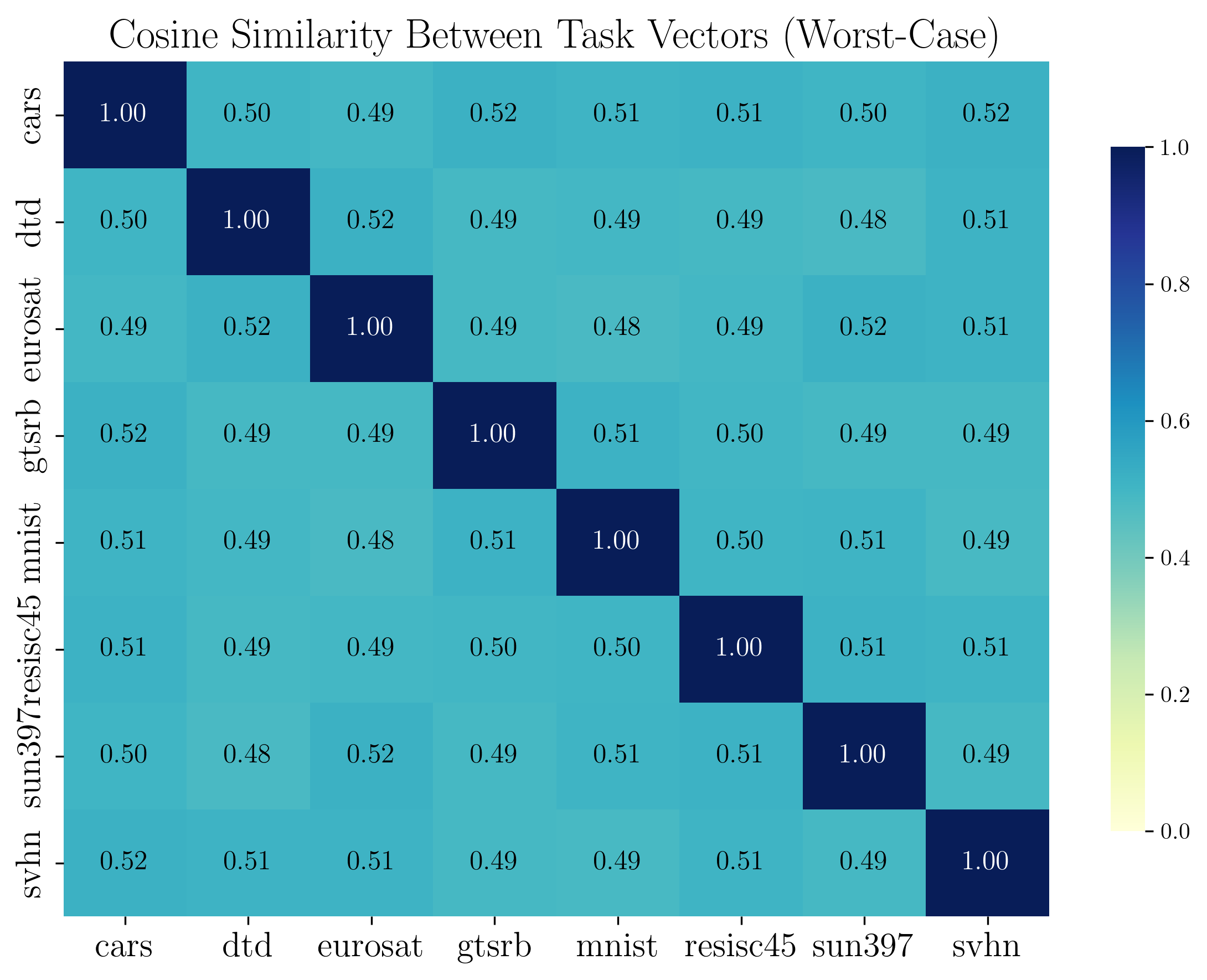}
    \caption{Cosine similarities between task vectors for the worst-case (SR). All $\hat{\tau}_q$ have increased similarity scores around 0.5, indicating less orthogonality, more weight entanglement, and increased cross-task interference.}
    \label{fig:mtllmf_wde_worst_case}
\end{figure}

To investigate the influence of wireless noise from a weight disentanglement perspective, we present the cosine similarities between task vectors as an alternative measure for WDE as in \cite{tang2024parameterefficientmultitaskmodel}. Figure \ref{fig:mtllmf_wde_RMTLLMF} shows results for R-MTLLMF (ideal case), and Figure \ref{fig:mtllmf_wde_worst_case} presents the adversarial worst-case (SR) without protection. A cosine similarity score of 0 indicates that the task vectors are orthogonal to each other and have no directional overlap. In contrast, a similarity of 1 indicates that all task vectors are perfectly aligned in the same direction and thus identical (same task). As can be seen, resilient R-MTLLMF protection helps to maintain small cosine similarities of at most 0.24, indicating that all task vectors are well disentangled and increasingly orthogonal to each other. This results in less cross-task interference and improved MTMF performance. In contrast, Figure \ref{fig:mtllmf_wde_worst_case} shows severe entanglement with cosine similarities between 0.48 and 0.52 across all tasks. This indicates that the task vectors are partially aligned, sharing increased directional similarity, which results in significant cross-task interference, reflected by the results in Figure \ref{fig:mtllmf_performance}. This also explains why performance remains detrimental regardless of the task count as entanglement is equally high for any considered combination.

\subsubsection{Ablation Studies}
\begin{figure}[!t]
    \centering
    \includegraphics[width=0.47\textwidth]{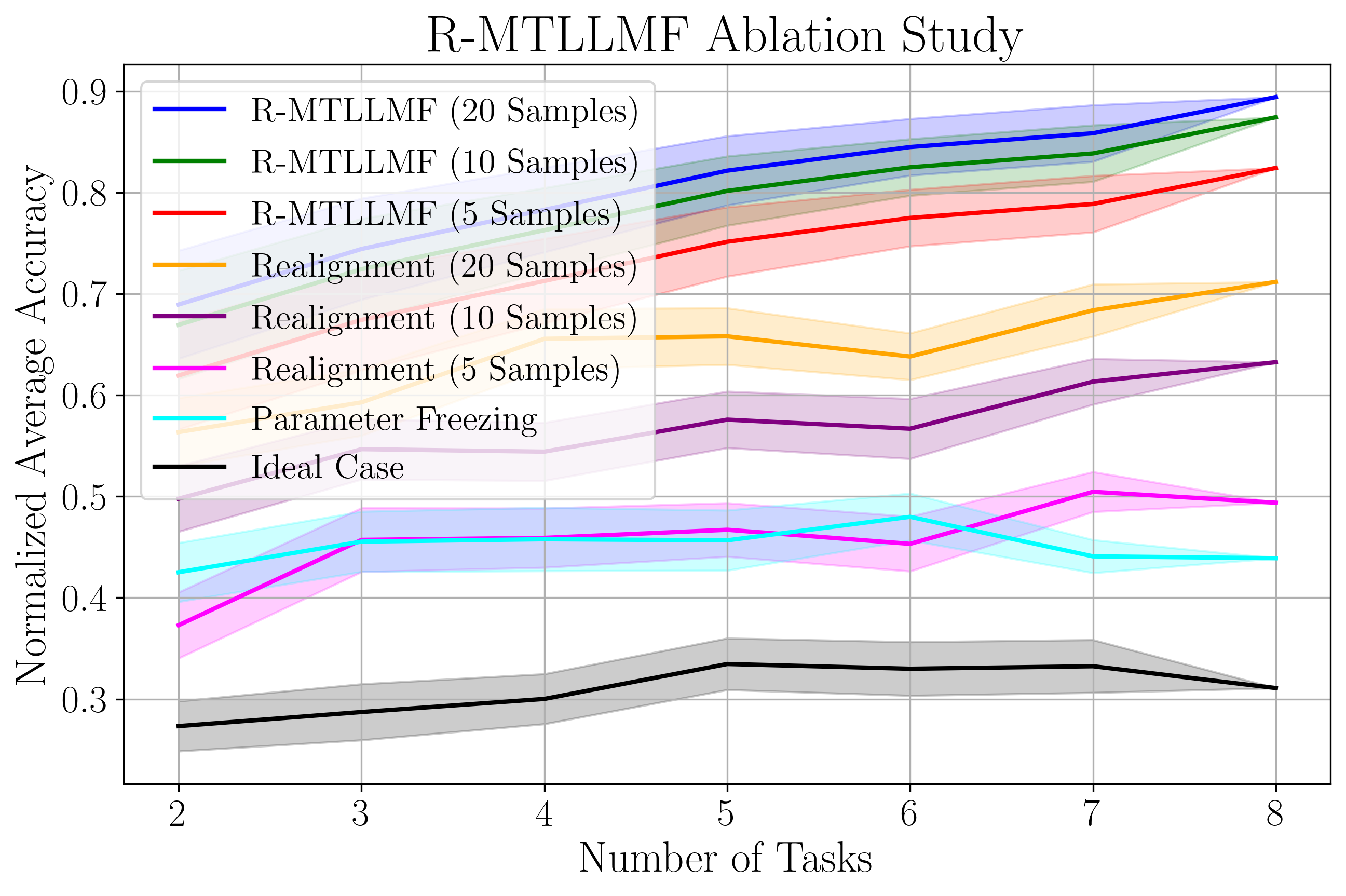}
    \caption{Ablation results for the ideal transmission case indicate that best performance is only achieved when all R-MTLLMF modules work together. Few-shot realignment with 10 samples is enough for satisfactory performance.}
    \label{fig:mtllmf_ablation}
\end{figure}

To understand the effectiveness of our R-MTLLMF framework better, we perform ablation studies in Figure \ref{fig:mtllmf_ablation} to study the importance of both parameter freezing and realignment, as well as the significance of the number of few-shot fine-tuning samples. Our results show that neither parameter freezing nor realignment alone can yield satisfactory performance, resulting in 21\% - 53\% worse average normalized accuracies as compared to full R-MTLLMF. We thus conjecture that alignment is more effective when essential embedding parameters are frozen, preserving the integrity of core data representations without the need for further optimization through fine-tuning. Furthermore, adding more samples to realignment generally increases performance, though with diminishing returns beyond 10 samples. This is especially true for the combined approach compared to realignment alone without parameter freezing. Therefore, both AI resilience strategies are necessary for R-MTLLMF's effectiveness.

In summary, MTMF improves the average performance as the number of tasks increases without the need for constant re-training. Our R-MTLLMF framework further provides an effective AI-based MTMF resilience framework for most typical wireless noise scenarios. It leads to increased task vector orthogonality and improves multi-task performance, achieving close-to-baseline results. However, its effectiveness diminishes

\vspace{-0.5cm}

\vspace{-0.5cm}

\newpage

\newpage
\noindent
under extreme adversarial noise. This suggests that AI-based resilience strategies alone may be insufficient for worst-case attacks, including sophisticated jamming that cannot be modeled as such and may lead to catastrophic denial of service \cite{boche_dos, boche_solv}. Thus, we recommend augmenting R-MTLLMF with complementary physical layer techniques, potentially enabled by future integrated sensing and communication (ISAC) \cite{djuhera2024rsfllmjammingresilientframework}. Despite this limitation, R-MTLLMF still outperforms unprotected setups by a factor of 9-14 under worst-case noise, making it a substantial improvement over standard MTMF.

%% file: extras/conclusion.tex
\section{Conclusion}
In this paper, we have studied the problem of adversarial noise attacks in MTMF at the wireless edge. To this end, we have first introduced MTMF via task vectors as a cost-effective alternative to construct MTLLMs without re-training. Second, we have derived a relationship between the cross-task interference and the MSE, showing a direct impact on performance. To mitigate this, we proposed R-MTLLMF, an AI-based framework for resilient task vector aggregation that employs few-shot fine-tuning and parameter freezing to realign the perturbed model. We have shown in extensive experiments that R-MTLLMF achieves close-to-baseline performance under typical wireless noise and effectively decreases cross-task interference. However, R-MTLLMF cannot fully compensate for worst-case adversarial noise as it does not ensure a resilient wireless link. We hypothesize that integrating physical layer protection schemes could further improve performance in extreme scenarios, creating holistic resilience from both wireless and AI perspectives. We leave this investigation for future work and advocate for further research into hybrid AI resilience mechanisms for distributed inference scenarios.